\documentclass[showpacs,preprintnumbers,amsmath,amssymb,preprint]{revtex4}
\usepackage{graphicx}
\usepackage{dcolumn}
\usepackage{bm}
\begin{document}
\title{Deuterium permeation in Er$_2$O$_3$ thin film fabricated on a type 316L stainless steel substrate}

\author{Halim \surname{Choi}}
\author{W. J. \surname{Byeon}}
\author{Yongmin \surname{Kim$^{*}$}\footnotetext[1]{ Author to whom correspondence should be addressed. Electronic mail : yongmin@dankook.ac.kr }}

\affiliation{Department of Physics, Dankook University, Cheonan 31116, Korea}

\begin{abstract}
  A metal-oxide film can be used as a hydrogen-isotope permeation barrier in the fuel circulation system for nuclear fusion. We fabricated Er$_2$O$_3$ thin film on a type 316L stainless-steel substrate by using a metal-organic chemical vapor deposition technique for the purpose of hydrogen-isotope permeation barrier. Electron microscopy based imaging and energy-dispersive X-ray spectroscopy measurements indicate a sound film quality together with X-ray diffraction experiments. We also measured deuterium permeation in the film at high temperatures ranging from 600 $^{\circ}$C to 800 $^{\circ}$C. The permeation reduction was most apparent at 650 $^{\circ}$C. Above 800 $^{\circ}$C, we confirmed that the film was damaged and did not work as a permeation barrier.
\end{abstract}



%\begin{keyword}
%Semiconductor, photoluminescence, diamagnetic shift
%\end{keyword}
%\end{frontmatter}
\maketitle

\section{Introduction}
Hydrogen isotopes of deuterium and tritium are fuels for nuclear fusion experiments. In the fusion fuel delivery and storage systems, such hydrogen-isotopes can permeate through the fuel lines and can be released  to  outside during the operation of a nuclear fusion reactor, which is an environmental issue. Coated metal-oxide (M-O) films  can reduce fuel permeation because the permeation process in M-O films is different from that in a bare stainless steel.  It is known that hydrogen easily permeates through SS fuel lines by thermal diffusion. However, when hydrogen atoms meet a surface of a M-O film, the hydrogen atoms tend to break M-O bond and form H-O bond until they replace all of the M-O bond\cite{Lee,Wu}.  Normally, the M-O bonding energy is larger than the H-O bonging energy. To elongate the permeation process, it is desired to have large bonding energy difference between the M-O and H-O bonds. Up to date, Al$_{2}$O$_{3}$ film coating was known to be an excellent permeation barrier. However, in the recent years, Er$_{2}$O$_{3}$ film has been suggested to be a good candidate \cite{Wu,Levchuk} for replacing Al$_{2}$O$_{3}$ film as a permeation barrier, because an Er$_{2}$O$_{3}$ film can be easily obtained and has a large M-O binding energy in comparison to an Al$_{2}$O$_{3}$ film. In this regard, various Er$_{2}$O$_{3}$  coating techniques and substrate materials are vastly adopted to improve hydrogen-isotope permeation reduction\cite{Wu,Levchuk,Yao,Liu,Chikada1,Mao,Chikada2,Wang,Li}. In the recent years, theoretical calculations on hydrogen permeation through grain boundaries on M-O films have been reported by using ab initio density fuctional theory, which indicated that proper analysis of the microstructures of M-O films is essential to understand of the hydrogen permeation mechanism\cite{Mao}.

In this study, we tried to deposit Er$_{2}$O$_{3}$ film on a type 316L stainless-steel (SS316L) substrate by using a chemical vapor deposition (CVD) technique. We obtain a 480-nm-thick Er$_{2}$O$_{3}$ film, which shows appreciable amount of permeation reduction factor (PRF) for deuterium at temperature below 700 $^{\circ}$C. With increasing temperature above 700 $^{\circ}$C, the PRF reduced drastically due to the fast activation of Er$_{2}$O3$_{3}$ film. We will discuss structural and permeation properties of coated Er$_{2}$O$_{3}$ film in the next sections.

\section{Experiment}

A schematic of the CVD system used for this study is shown in Fig. 1. A 50-mm-diameter and 600-mm-long quartz tube was used as the main reactor chamber. The quartz reactor is surrounded by a heater that can heat the reactor up to 1200 $^{\circ}$C. A 20-mm-diameter with 1-mm-thick SS316L coin shaped substrate was located on a square-shape quartz substrate holder, which is equipped at the reactor center with a vertical angleof 30$^{\circ}$. The substrates were carefully polished to a mirror finish by using a lapping machine. Before being coated, the substrates were heat treated at 700 $^{\circ}$C in the reactor in hydrogen atmosphere to removed the surface contaminant. Tris(2,2,6,6-tetramethyl-3,5-heptanedionato) erbium, or simply called Er(tmhd)3 was used as a precursor for Er$_{2}$O$_{3}$ coating. The Er(tmhd)3 precursor is known as stable while transporting from the sublimator, which is quickly decomposed and formed to Er$_{2}$O$_{3}$ above 500 $^{\circ}$C in oxygen atmosphere. The sublimator and the transport tubes are composed of stainless steel, which were wrapped with heater stripes. The whole units including sublimator, transport lines and heater stripes were wound by thermal insulating materials to reduce the temperature fluctuation during the process.  High purity Ar gas was used as a precursor carrier, which is regulated by mass-flow controller (MFC) with the flow rate of 50 sccm. After passing through the MFC, Ar gas was heated to 180 $^{\circ}$C. Therefore, the sublimated precursor's temperature was maintained at 180 $^{\circ}$C in the sublimator and the transport lines. The flow rate of O$_{2}$ gas was 100 sccm. During the coating process, the pressure inside of the reactor was maintained at 10 mbar by using an automatic pressure regulating valve, which is mounted between the chamber and the vacuum pump.  The total deposition time was 4 hrs. and after finishing the film deposition process, samples were slowly cooled down by 0.5 $^{\circ}$C/min.  to room temperature. One thing of note is the clogging of the precursor in the precursor transport line after the film growth. Even though Er(tmhd)3 is known as a stable precursor, a significant amount of the precursor was deposited inside of the transport line. Such precursor clogging in the transport line not only elongated the coating time but also deteriorated the sample quality. Therefore, the used transport line had to be replaced with a new one after every deposition.

	The estimated Er$_2$O$_3$ film thickness was measured to be 395 nm by using an ellipsometer (Woolim Model M-2000). For further analysis of the coated Er$_2$O$_3$ film, we performed scanning electron microscopy (SEM) image, energy dispersive spectrometry (EDS), and X-ray diffraction measurements. To obtain the deuterium permeation, we used a home-built hydrogen-isotope permeation measurement system (HPMS). Figure 1b shows a schematic diagram of the HPMS, which consists of a sample mount, a heater and residual gas analysis (RGA) modules. The supplying deuterium gas pressure was maintained at 1 bar during the permeation measurements. To minimize the sample temperature fluctuation, the heater unit is located in a vacuum chamber as a thermal shield.
The partial pressure of deuterium at the permeate side was monitored by using the RGA, which is converted to the permeability of deuterium. The detailed development and operation procedure of the HPMS can be found elsewhere\cite{Lee2,Lee3}.

\section{Results and Discussion}
	By using an SEM and an EDS, we obtained surface images and atomic composition of the sample before and after the permeation measurements. Figure 2a exhibits an SEM image of a fresh  as-grown Er$_{2}$O$_{3}$ film before the permeation measurements. It shows granular shape grains with the average size of 35 nm. After the high temperature permeation measurements at 850 $^{\circ}$C, as seen in Fig. 2b, the average grain size was increased to $\sim$ 50 nm. However, the film was damaged and no longer worked as a permeation barrier at this elevated temperature.

The EDS analysis shows that the atomic percentile contents of O and Er is 43.8 \% and 22.7 \%, respectively, for as-grown sample before the permeation experiments. The residual 33.5 \% comes from the substrate (SS316L).  The content ratio between O to Er is 1.93, which is higher than the stoichiometric ratio 1.5 due to the surplus oxygen contents.  After the permeation experiments, the atomic percents of O and Er changed to 34.1 \% and 17.0 \%, respectively. In this case, the atomic content ratio between O and Er is 2.0. Exposure to high temperature deuterium may reduce the amount of O and Er contents. The similar contents ratio before and after the permeation measurements suggests that the surplus oxygen contents may come from the native oxide formed on the SS316L surface. Even though the sample exposed to the high temperature deuterium, it cannot affect the native oxide on the surface of SS316L substrate, which does not alter the O to Er contents ratio.

	The XRD spectra of Er$_{2}$O$_{3}$ and SS316L substrate are displayed in Fig. 3. The peaks at 50.7$^{\circ}$ and 74.6$^{\circ}$ are from FeCrNiC SS316L substrate. The peak at 43.6$^{\circ}$ is the combined peak of Er$_{2}$O$_{3}$ (431) and SS316L (111) planes, and all other peaks are indexed to be of Er$_{2}$O$_{3}$. As seen in the figure, the XRD peak intensities of Er$_{2}$O$_{3}$ became sharper (smaller line-widths) and stronger after the permeation experiments compared with those of the as-grown sample. This is due to the improvement of the crystallinity by the long-term exposure at high-temperature during the permeation experiments. However, though the improvement of the crystallinity, the Er$_{2}$O$_{3}$ film was significantly damaged due to the large mismatch of the thermal expansion coefficients between the substrate and the coated film  at such high temperature as seen in Fig. 2b.

Arrhenius  plots of the deuterium permeabilities for both the base and Er$_{2}$O$_{3}$-coated SS316L samples are displayed in Fig. 4.  Solid square markers are the data measured in this study. To compare our permeability data with those of the similar deposition technique from other research group, data extracted from Ref. \cite{Wu} are displayed as circular markers. The permeability was recorded when deuterium permeation reached the steady state. We measured permeability at five different temperature points from 600 $^{\circ}$C to 800 $^{\circ}$C by 50 $^{\circ}$C steps.  As seen in the figure, the permeability increases with increasing temperature. Below 700 $^{\circ}$C, the slope and the amount of the permeability is moderate and small, respectively. However, above 700 $^{\circ}$C, the permeability slope rapidly increases. In comparison, the  values measured in this study are significantly lower than the values from Ref. \cite{Wu} (solid circle). Permeation reduction factor (PRF), which is the ratio between the permeability of SS316L and Er$_{2}$O$_{3}$, is summarized in Table 1. A larger PRF value indicates a better permeation protection. The PRF at 600 $^{\circ}$C is 592, which increases to 881 at 650 $^{\circ}$C, then decreases to 510 at 700 $^{\circ}$C. It is not clear such a sudden increase of PRF at 650 $^{\circ}$C. One plausible explanation can be an annealing effect. As mentioned above, the permeability values were recorded at the steady state; it takes hours to days to be reached at the steady state depending on the designated temperature. During the 600 $^{\circ}$C and 650 $^{\circ}$C measurements, the sample was exposed several hours at such elevating temperature exceeding the growth temperature, and the improved crystallinity of the sample during the measurements possibly increases the PRF at 650 $^{\circ}$C.

%High temperature region above 700 $^{\circ}$, H-O bonding activates inside of the film quickly such that the permeability slop rapidly increased for which the RPF reduced to 23 at 800  $^{\circ}$. We attempted to measure permeation at 850  $^{\circ}$. However, due not only to the high activation, but also to the cracks on the film as seen in Fig. 2 (b), the film did not play as a permeation barrier.

\begin{table*}
\caption{\label{tab:table1}Permeation reduction factors at different temperatures}
\begin{ruledtabular}
\begin{tabular}{ccccccc}
 temp. $^{\circ}$C & 600 & 650 & 700 & 750 & 800   \\
\hline
from Ref. \cite{Wu} & 291 & 235 & 151 & - & - \\%\footnotemark[1] \\
this study & 592 & 881 & 510 & 76 & 23  \\% \footnotemark[2] \\

\end{tabular}
\end{ruledtabular}
%\footnotetext[1]{Here's the first, from Ref.~\onlinecite{feyn54}.}
%\footnotetext[2]{Here's the second.}
%\footnotetext[3]{Here's the third.}
\end{table*}

\section{Conclusion}

We fabricated an Er$_{2}$O$_{3}$ film on a SS316L substrate by using a CVD method. After 4-hour growth, the film thickness was measured to be 395 nm by using an ellipsometry.  SEM images indicate that grains with average size of $\sim$ 30 nm for an as-grown polycrystal film become large enough to induce crack damages after the deuterium permeation measurements at 850 $^{\circ}$C. By using the home-built HPMS, we measured the deuterium permeability  of Er$_{2}$O$_{3}$ film at temperatures between 600 $^{\circ}$C and 800  $^{\circ}$C by 50  $^{\circ}$C steps. The measured permeation reduction was most apparent at 650 $^{\circ}$C. Above 700 $^{\circ}$C, the permeability decreases rapidly with increasing temperature, and at 850 $^{\circ}$C, the film was damaged and no longer worked as a permeation barrier.

\section*{Acknowledgments}
The present research was supported by the research fund of Dankook University in 2017. One of the author, YK thanks Dr. Y. H. Shin and Prof. S. J. Noh for their experimental supports.
%This work was supported by the "ITER Technology R\&D Program(IN1807-4)" through the National Fusion Research Institute (NFRI) of Korea funded by the Korean Government, and by the National Research Foundation of Korea (Project No. 2018R1A2B6002797)

\section*{References}
%\bibliographystyle{model1a-num-names}

%\begin{thebibliography}{00}

%\end{thebibliography}

\newpage

\begin{figure}
\caption{Fig 1 (a) A schematic drawing of the CVD system used for this study, in which precursor and O$_{2}$ flows, temperature and base pressure of the grow chamber were precisely controlled. (b) A schematic drawing of the home-built Hydrogen Permeation Measurement System (HIPMS). To avoid temperature fluctuation during the permeation measurements, the heating system is surrounded by a large volume vacuum thermal barrier.  }
\label{fig1}
\end{figure}

\begin{figure}
\caption{Fig. 2 (a) SEM image of the as-grown sample before the permeation measurements. It shows granular-shape polycrystals with the average grain size of $\sim$ 30 nm. (b) SEM image of the sample after the permeation measurements. Due to high temperature exposure, the film is shown to be damaged.   }
\label{fig2}
\end{figure}

\begin{figure}
\caption{Fig. 3 X-ray diffraction peaks of SS316L bare substrate (bottom) and as-grown Er$_{2}$O$_{3}$ film on a SS316L substrate before and after the permeation experiments.    }
\label{fig3}
\end{figure}

\begin{figure}
\caption{Fig. 4 Arrhenius plot of deuterium permeability. Red solid line, blue square, red circular markers are permeabilities of SS316L, our data and from Ref. \cite{Wu}, respectively. The permeability of our data is significantly lower than that of Ref. 10 below 700 $^{\circ}$C. Above 700 $^{\circ}$C, the permeability rapidly increased.     }
\label{fig5}
\end{figure}


\begin{references}

\bibitem{Lee} R. W. Lee , R. C. Frank and D. E. Swets, J. Chem. Phys. \textbf{36}, 1062 (1962).

\bibitem{Wu} Y. Wu, D. He, S. Li, Y. Yang, X. Liu, S. Wang and L. Jiang, Inter. J. of Hydrogen Energy \textbf{41}, 431  (2016).

\bibitem{Levchuk} D. Levchuk, S. Levchuk, H. Maier, H. Bolt and A. Suzuki, J. Nucl. Mater. \textbf{367–370}, 1033 (2007).

\bibitem{Yao} Z. Yao, A. Suzuki, D. Levchuk, T. Chikada, T. Tanaka and T. Muroga, Journal of Nuclear Materials \textbf{386–388}, 700  (2009).

\bibitem{Liu} S. Liu, X. Ju, Y. Xin, J. Qiu, T. Li and J.-L. Cao, Fusion Engineering and Design \textbf{85}, 1401  (2010).

\bibitem{Chikada1} T. Chikada, S. Naitoh, A. Suzuki, T. Terai, T. Tanaka and T. Muroga, Journal of Nuclear Materials \textbf{442}, 533  (2013).

\bibitem{Mao} W. Mao, T. Chikada, K. Shimura, A. Suzuki, K. Yamaguchi and T. Terai, Journal of Nuclear Materials \textbf{443}, 555  (2013).

\bibitem{Chikada2} T. Chikada, M. Shimada, R. J. Pawelko, T. Terai and T. Muroga, Fusion Engineering and Design \textbf{89}, 1402  (2014).


\bibitem{Wang} J. Wang, Q. Li, Q.-Y. Xiang, T. Tang, Y.-C. Rao and J.-L. Cao, Inter. J. of Hydrogen Energy \textbf{41}, 1326  (2016).


\bibitem{Li} Q. Li, J. Wang, Q. Xiang, K. Yan, T. Tang, Y.-C. Rao and J.-L. Cao, Journal of the European Ceramic Society \textbf{37}, 249  (2017).


\bibitem{Lee2}	S. K. Lee, S-H. Yun, H. G. Joo and S. J. Noh, Curr. Appl. Phys. \textbf{14}, 1385 (2014).

\bibitem{Lee3}	H. S. Kim, W. J. Byun and S. J. Noh, J. Kor. Phys. Soc. \textbf{72}, 1 (2018).


%\bibitem{Verghese} K. Verghese, L.R. Zumwalt, C.P. Feng, T.S. Elleman, J. Nucl. Mater. \textbf{85-86}, 1161 (1979).

%\bibitem{Forcey} K.S. Forcey, D.K. Ross, C.H. Wu, J. Nucl. Mater. \textbf{182},  36 (1991).

%\bibitem{Song} R.G. Song, Surf. Coat. Technol. \textbf{168},  191 (2003).

%\bibitem{Hollenberg} G.W. Hollenberg, E.P. Simonen, G. Kalinin, A. Terlain, Fus. Eng. Des. \textbf{28}, 190 (1995).



\end{references}
\end{document}